\documentclass[aps,prb,twocolumn]{revtex4-2}
\usepackage{tabularx} 
\usepackage{amsmath}  
\usepackage{graphicx} 
\usepackage[final]{hyperref} 
\usepackage{multirow}
\hypersetup{
	colorlinks=true,       
	linkcolor=blue,        
	citecolor=blue,        
	filecolor=magenta,     
	urlcolor=blue         
}
\usepackage{blindtext}
\usepackage{verbatim}
\usepackage{siunitx}

\begin{document}

\title{Exchange-driven spin Hall effect in anisotropic ferromagnets}
\author{K. D. Belashchenko}
\affiliation{Department of Physics and Astronomy and Nebraska Center for Materials and Nanoscience, University of Nebraska-Lincoln, Lincoln, Nebraska 68588, USA}
\date{\today}

\begin{abstract}
Crystallographic anisotropy of the spin-dependent conductivity tensor can be exploited to generate transverse spin-polarized current in a ferromagnetic film. This ferromagnetic spin Hall effect is analogous to the spin-splitting effect in altermagnets and does not require spin-orbit coupling. First-principles screening of 41 non-cubic ferromagnets revealed that many of them, when grown as a single crystal with tilted crystallographic axes, can exhibit large spin Hall angles comparable with the best available spin-orbit-driven spin Hall sources. Macroscopic spin Hall effect is possible for uniformly magnetized ferromagnetic films grown on some low-symmetry substrates with epitaxial relations that prevent cancellation of contributions from different orientation domains. Macroscopic response is also possible for any substrate if magnetocrystalline anisotropy is strong enough to lock the magnetization to the crystallographic axes in different orientation domains.
\end{abstract} 

\maketitle

\section{Introduction}

The spin Hall effect \cite{DYAKONOV1971459,Sinova-SHE} is utilized for generating transverse spin currents which can be injected across interfaces into adjacent layers in heterostructures. If that spin current is injected into a magnetic layer, it can be converted into magnetization torque. For heavy-metal spin Hall layers, this is called spin-orbit torque \cite{Manchon_RMP} and is used as an alternative to spin-transfer torque for applications including magnetic random-access memory (MRAM), high-frequency oscillators, and neuromorphic computing \cite{Roadmap,Song2021}.

The spin Hall effect also exists in ferromagnets \cite{Miao2013,Wang2014,Seemann2015,Taniguchi2015,Tian2016,Wu2017,Das2017,Gibbons2018,Bose2018,Omori2019,Amin2019,Davidson2020,Salemi2022}, where it has both time-reversal even and odd \cite{Mook2020,Salemi2022} components. The odd component, referred to as the magnetic spin Hall effect, was also observed in noncollinear antiferromagnets \cite{Zelezny2017,Kimata2019,Holanda2020,Kondou2021,Hu2022,Han2023}. The structure of the spin current response tensor in the presence of spin-orbit coupling may be deduced from the magnetic point group of the material \cite{Seemann2015}.
Some magnets can exhibit spin Hall response even without spin-orbit coupling, in which case the symmetry of the response tensor may be determined using spin groups treating spatial and spin degrees of freedom separately from each other \cite{Litvin1974}. This was first noted for noncollinear antiferromagnets \cite{Zelezny2017,Kimata2019}.

Transverse spin currents without spin-orbit coupling are also possible \cite{Gonzalez2021,Bai2022,Karube2022,Bose2022} in so-called altermagnets 
\cite{Smejkal2022,Smejkal2022landscape}, i.e., collinear antiferromagnets with $k$-dependent exchange splitting allowed by symmetry. In this case, spin-orbit coupling usually plays a minor role and can be neglected, but the crystallographic orientation of the film is important. The exchange-driven spin Hall current in altermagnets has been referred to as the \emph{spin-splitting current} \cite{Gonzalez2021,Bai2022,Karube2022}. 
For collinear two-sublattice antiferromagnets without spin-orbit coupling, the symmetry of the macroscopic response properties, including the spin Hall tensor, is fully described by the black-and-white (Heesch-Shubnikov) point group, while the polarization of the spin current is decoupled from the crystal axes and is simply parallel to the antiferromagnetic order parameter \cite{Turek}.

In contrast to the spin Hall effect in heavy metals, which is typically dominated by the intrinsic mechanism \cite{Sinova-SHE}, the spin-splitting current in altermagnets scales with the longitudinal conductivity, leading to a disorder-independent spin Hall angle. This feature is shared with the magnetic spin Hall effect, which is also time-reversal-odd \cite{Salemi2022}. The predicted spin Hall angle for altermagnetic RuO$_2$ is about 0.27 \cite{Gonzalez2021}, which compares favorably with spin-orbit-driven sources like Pt or $\beta$-W. However, observation of this large response may be hindered by the difficulty in achieving a single-domain state in RuO$_2$ due to the absence of uncompensated magnetization \cite{Karube2022}.

In this paper, I show that exchange-driven spin Hall effect can also be observed in ferromagnets and ferrimagnets thanks to the anisotropy of the spin-dependent conductivity tensor. As in the case of altermagnets, generation of the transverse spin current imposes restrictions on the bulk symmetry and the orientation of the film. In particular, the ferromagnet can not be cubic, and it needs to be grown such that the surface normal does not coincide with a principal axis of the conductivity tensor. Using first-principles calculations, I then estimate the spin Hall angles for a number of tetragonal, hexagonal, and orthorhombic ferromagnets and ferrimagnets, identifying a few with spin Hall angles as large as 0.3--0.4. Where possible, suitable substrates are suggested for growing films with the tilted high-symmetry axes, as required for the observation of the exchange-driven spin Hall current.
If the film is uniformly magnetized, observation of the spin Hall current over a macroscopically large surface area requires that the bilayer has sufficiently low symmetry so that the spin Hall current is not cancelled out by averaging over degenerate orientation domains. These symmetry requirements may be fully relaxed if the ferromagnet has strong magnetocrystalline anisotropy which locks the magnetization to the crystallographic axes in different orientation domains.

The paper is organized as follows. Section \ref{sec:theory} presents the analysis of the exchange-driven spin Hall response in a film with an anisotropic spin-dependent conductivity tensor. Section \ref{sec:computational} describes computational details, and the results are displayed in Section \ref{sec:results}. Section \ref{sec:domains} discusses crystallographic orientation domains and the statistical average of the spin Hall conductivity over these domains. The conclusions are summarized in Section \ref{sec:conclusions}.

\section{Transverse spin current in a magnetic film with open boundary conditions}
\label{sec:theory}

Consider a metallic film with a collinear magnetization whose spin-dependent conductivity tensor is $\sigma^\lambda_{\alpha\beta}$, where the lower indices denote Cartesian components, and $\lambda\in\{\uparrow,\downarrow\}$ is the spin projection. I will also use $\lambda=\pm1$ in equations. The conductivity tensor is symmetric because spin-orbit coupling is neglected, which imposes time-reversal invariance independently in each spin channel, and the spin projection is defined with respect to the orientation of the magnetic order parameter. Let the $z$ axis point along the film normal direction and the $x$ axis along the externally applied electric field $E_x$.

Because $\sigma^\lambda_{\alpha\beta}$ is anisotropic, the charge Hall response will generally lead to the build-up of Hall voltage across the thickness of the film. The resulting electric field component $E_z$ is determined under the assumption of open-circuit boundary conditions, i.e., vanishing charge current density perpendicular to the film plane, $j_z=0$. We then find the steady-state bulk spin current flowing perpendicular to the film plane. This spin current corresponds to the bulk spin Hall source term. To determine the spin current and accumulation profiles in a specific multilayer, this source term needs to be included in the spin-diffusion equations \cite{ValetFert,ZuticRMP2004}. A somewhat similar problem was considered for the transverse spin current driven by relativistic sources in a ferromagnetic film \cite{Taniguchi2015}.

Explicitly, we have
\begin{align}
    j_x &= \sum_\lambda\left[\sigma^\lambda_{xx}E_x + \sigma^\lambda_{xz}E_z\right]\,\\
    j_z &= \sum_\lambda\left[\sigma^\lambda_{zx}E_x + \sigma^\lambda_{zz}E_z\right]\,\\
    j^s_z &= \sum_\lambda\left[\sigma^\lambda_{zx}E_x + \sigma^\lambda_{zz}E_z\right]\lambda\,
\end{align}
where $j^s_z$ is the spin current flowing in the $z$ direction. Setting $j_z=0$ and eliminating $E_z$, we find the effective spin Hall conductivity $\sigma_\mathrm{SH}=j^s_z/E_x$ and the effective longitudinal conductivity $\tilde\sigma_{xx}=j_x/E_x$ under open boundary conditions:
\begin{align}
    \sigma_\mathrm{SH} &= \frac{\sigma_{zz}\sigma^s_{zx}-\sigma^s_{zz}\sigma_{zx}}{\sigma_{zz}}\label{sigmaSH}\,\\
    \tilde\sigma_{xx} &= \frac{\sigma_{xx}\sigma_{zz}-\sigma^2_{xz}}{\sigma_{zz}}\label{sigmaxx}
\end{align}
where $\sigma_{\alpha\beta}=\sum_\lambda\sigma^\lambda_{\alpha\beta}$ and $\sigma^s_{\alpha\beta}=\sum_\lambda\lambda\,\sigma^\lambda_{\alpha\beta}$. We are interested in the spin Hall angle under open boundary conditions defined as $\theta_\mathrm{SH}=\sigma_\mathrm{SH}/\tilde\sigma_{xx}$.

For the given crystallographic orientation of the film, the charge and spin conductivity tensors in Eqs. (\ref{sigmaSH})-(\ref{sigmaxx}) may be expressed through their principal components. For simplicity, let us assume the principal axes of $\sigma_{\alpha\beta}$ and $\sigma^s_{\alpha\beta}$ are the same, which is true if they are fixed by symmetry; this is the case for tetragonal, hexagonal, and orthorhombic lattices considered below. Further, let us also assume that one of the principal axes is oriented in the plane of the film along the $y$ axis. I denote the other two principal axes as 1 and 2 and the angle made by principal axis 1 with the $z$ axis as $\theta$. Applying the rotation by angle $\theta$ with respect to the $y$ axis to the tensors expressed in their principal axes and substituting in Eqs. (\ref{sigmaSH})-(\ref{sigmaxx}), we find, after some algebra:
\begin{equation}
    \theta_\mathrm{SH}=\frac12(\beta_1-\beta_2)\sin2\theta
    \label{SHA}
\end{equation}
where $\beta_i=\sigma^s_i/\sigma_i$ is the conventionally defined \cite{ZuticRMP2004,Bass2016} spin polarization of the conductivity along its principal axis $i$.

Equation (\ref{SHA}) shows that $|\theta_\mathrm{SH}|\leq 1$, and the maximal value of 1 is reached if $\theta=\pi/4$ and $\beta_1=-\beta_2=1$. This maximal condition requires that the current along principal axes 1 and 2 is carried exclusively by spin-up and spin-down electrons, respectively, i.e., the ferromagnet behaves as a half-metal with opposite spin polarizations along the two principal axes. While this condition is unlikely to be satisfied by any material, we will see below that values of $|\theta_\mathrm{SH}|$ exceeding 0.1 are quite common, which compares favorably with the spin Hall angles typical of heavy metals such as Pt \cite{Sinova-SHE}.

In the relaxation-time approximation \cite{ziman1960book}, the conductivity tensor is given by
\begin{equation}
\sigma^\lambda_{\alpha\beta}=e^2\sum_n\int v^\lambda_{n\alpha} v^\lambda_{n\beta}\tau^\lambda_n\frac{\partial f(E^\lambda_n)}{\partial \mu}\frac{d^3k}{(2\pi)^3}
\label{RTA}
\end{equation}
where $\mathbf{v}^\lambda_n(\mathbf{k})$, $\tau^\lambda_n(\mathbf{k})$, and $E^\lambda_n(\mathbf{k})$ are the group velocity, relaxation time, and energy of the Bloch band $n$ of spin $\lambda$, and $f(E)$ is the Fermi-Dirac distribution function at chemical potential $\mu$ for the given temperature. The relaxation-time approximation (\ref{RTA}) is appropriate if disorder-induced broadening does not lead to significant overlap between electronic bands \cite{Tsymbal1996}.

The relaxation times generally depend on the kind of impurities or other disorder present in the system, but, for a typical good metal with an extended Fermi surface and localized disorder that is typical for sputtered multilayers, it is reasonable to assume that $\tau^\lambda_n(\mathbf{k})$ depends only on spin $\lambda$. Then $\tau_\lambda$ can be pulled out of the integral in (\ref{RTA}), and we have $\sigma^\lambda_{\alpha\beta}=e^2\tau_\lambda K^\lambda_{\alpha\beta}$, where the tensor $K^\lambda_{\alpha\beta}$ depends only on the electronic structure and can be calculated using standard techniques. 

Introducing the spin polarizations of the relaxation time, $P_\tau=(\tau_\uparrow-\tau_\downarrow)/(\tau_\uparrow+\tau_\downarrow)$, and of the principal value $i$ of tensor $K$, $P_i=(K^\uparrow_i-K^\downarrow_i)/(K^\uparrow_i+K^\downarrow_i)$, we find
\begin{align}
    \beta_i = \frac{P_\tau + P_i}{1+P_\tau P_i}
\end{align}
and
\begin{align}
    \beta_1-\beta_2 = \frac{(1-P^2_\tau)(P_1-P_2)}{(1+P_1P_\tau)(1+P_2P_\tau)}.
\label{b1b2}
\end{align}

Equation (\ref{b1b2}) shows that the difference $P_1-P_2$, which is an easily calculable band structure property, can be used as a descriptor to screen for materials with large $|\beta_1-\beta_2|$. We should, however, bear in mind that this descriptor is only adequate as long as $P_\tau$ is not too large, in which case $|\beta_1-\beta_2|$ can be significantly reduced compared to $|P_1-P_2|$. On the other hand, if $P_1\neq -P_2$, there is a range of $P_\tau$ of a certain sign that makes $|\beta_1-\beta_2|$ greater than $|P_1-P_2|$. 

The (isotropic) transport spin polarization $\beta$ in elemental ferromagnets Fe, Co, and Ni alloyed with a small amount of a second magnetic element varies drastically, including in sign, depending on the choice of that element \cite{Campbell1982,Bass2016}. This suggests that magnetic impurities can be used to tune $P_\tau$ in a wide range as long as scattering is dominated by these impurities.

The spin polarization $P$ of the isotropic tensor $K$ in bcc Fe, fcc Co, and fcc Ni, calculated as described in Section \ref{sec:computational}, is, respectively, 0.34, 0.05, and 0.02. The values for Co and Ni are small compared to the typical measured values of $\beta$ in multilayer structures, such as 0.3--0.5 in Co or 0.5-0.8 in permalloy \cite{Bass2016}. A theoretical calculation with explicit treatment of substitutional disorder in fcc Ni$_{1-x}$Fe$_x$ alloys resulted in $\beta$ exceeding 0.5 in the whole compositional range \cite{Starikov2010}. This comparison suggests that $\beta$ in these ferromagnets is often dominated by $P_\tau$. However, in searching for materials with a large \emph{anisotropy} of $\beta$, which, according to (\ref{SHA}), is required for efficient transverse spin current generation, the anisotropy of $P$ is of central importance, according to (\ref{b1b2}). Therefore, in the following I use the quantity $\tilde\theta_{ij}=\frac12(P_i-P_j)$ as a proxy for the spin Hall angle of a film with geometry described before Eq. (\ref{SHA}). 

\section{Computational details}
\label{sec:computational}

Using the Materials Project database \cite{MaterialsProject}, a set of experimentally reported ferro- and ferrimagnetic materials was selected using the following criteria: no more than three constituent elements, no rare earths (with the exception of GdNi) or actinides, metallic and magnetic DFT ground state, at least one $3d$ atom from V to Ni, with the tetragonal, hexagonal, or orthorhombic crystal structure. This set was further pruned, removing materials with no experimentally reported magnetic phase transition, in part using the database supplied with Ref. \onlinecite{CourtCole2018}. Half-metallic, quasi-two-dimensional, and weakly ferromagnetic materials were also excluded, along with those with a large number of atoms per unit cell. The resulting set is by no means complete, but it provides a broad, representative selection.

The lattice structures were taken from the Materials Project database \cite{MaterialsProject} and standardized using the \texttt{spglib} library \cite{spglib} with a large tolerance of 0.5\% to remove the occasional spurious symmetry reductions. The structures were then optimized using the projector-augmented wave (PAW) method \cite{BLOCHL1994}, implemented in the Vienna Ab Initio Simulation Package (VASP) \cite{VASP1,VASP2,VASP3}, using the generalized gradient approximation of Perdew-Burke-Ernzerhof (PBE) \cite{PBE} for the exchange-correlation functional. The optimized lattice constants and magnetizations are listed in Table \ref{tab:latcon}.

For ferrimagnetic Mn$_2$Sb, it was found that the LDA$+U$ method with $U=2.0$ eV applied to the $3d$ shell of the Mn atoms in octahedral sites significantly improves agreement with experimental data ($a=4.03$ \AA, $c=6.53$ \AA, $M=1.66$ $\mu_B/$f.u.) \cite{Mn2Sb_experiment} for the lattice constants and the magnetization. This calculation is labeled Mn$_2$Sb$+U$ in Table \ref{tab:latcon}.

\begin{table*}[htb]
\caption{Optimized ($a$, $b$, $c$) and experimental \cite{CSD} ($a_\mathrm{exp}$, $b_\mathrm{exp}$, $c_\mathrm{exp}$) lattice constants (\AA) and calculated magnetization $M$ ($\mu_B$/f.u.) for selected materials. Tetragonal, hexagonal, and orthorhombic magnets are separated by horizontal lines. MPID is the Materials Project identifier \cite{MaterialsProject}. Mn$_2$Sb$+U$ is Mn$_2$Sb treated with LDA$+U$ (see text).}
\centering
\begin{tabular}{|l|l|c|c|c|c|c|c|c|}
\hline
Material & MPID & $a$ & $b$ & $c$ & $a_\mathrm{exp}$ & $b_\mathrm{exp}$ & $c_\mathrm{exp}$ & $M$ \\
\hline
FePt & mp-2260 & \multicolumn{2}{c|}{2.723} & 3.764 & \multicolumn{2}{c|}{2.729} & 3.713 & 3.28\\ 
FePd & mp-2831 & \multicolumn{2}{c|}{2.708} & 3.770 & \multicolumn{2}{c|}{2.722} & 3.714 & 3.29\\ 
CoPt & mp-949 & \multicolumn{2}{c|}{2.690} & 3.715 & \multicolumn{2}{c|}{2.691} & 3.684 & 2.27\\ 
MnAl & mp-771 & \multicolumn{2}{c|}{2.755} & 3.476 & \multicolumn{2}{c|}{2.772} & 3.57 & 2.32\\ 
MnGa & mp-1001836 & \multicolumn{2}{c|}{2.705} & 3.660 & \multicolumn{2}{c|}{2.748} & 3.676 & 2.51\\ 
FeNi & mp-2213 & \multicolumn{2}{c|}{2.514} & 3.577 & \multicolumn{2}{c|}{2.533} & 3.582 & 3.24\\ 
Fe$_2$B & mp-1915 & \multicolumn{2}{c|}{5.046} & 4.228 & \multicolumn{2}{c|}{5.11} & 4.249 & 3.67\\ 
Co$_2$B & mp-493 & \multicolumn{2}{c|}{4.947} & 4.245 & \multicolumn{2}{c|}{5.015} & 4.22 & 2.00\\ 
MnAu$_4$ & mp-12565 & \multicolumn{2}{c|}{6.520} & 4.045 & \multicolumn{2}{c|}{6.45} & 4.04 & 4.16\\ 
VAu$_4$ & mp-1069697 & \multicolumn{2}{c|}{6.507} & 4.035 & \multicolumn{2}{c|}{6.4} & 3.98 & 1.95\\ 
Mn$_2$Sb & mp-20664 & \multicolumn{2}{c|}{3.921} & 6.419 & \multicolumn{2}{c|}{4.085} & 6.534 & 1.28\\ 
Mn$_2$Sb$+U$ & mp-20664 & \multicolumn{2}{c|}{4.026} & 6.527 & \multicolumn{2}{c|}{4.085} & 6.534 & 1.60\\ 
Fe$_3$B & mp-1181327 & \multicolumn{2}{c|}{8.535} & 4.227 & \multicolumn{2}{c|}{8.647} & 4.282 & 5.82\\ 
Fe$_3$P & mp-18708 & \multicolumn{2}{c|}{9.042} & 4.374 & \multicolumn{2}{c|}{9.107} & 4.460 & 5.68\\ 
Fe$_8$N & mp-555 & \multicolumn{2}{c|}{5.681} & 6.223 & \multicolumn{2}{c|}{5.718} & 6.288 & 19.3\\ 
Mn$_2$Ga$_5$ & mp-607225 & \multicolumn{2}{c|}{8.809} & 2.692 & \multicolumn{2}{c|}{8.803} & 2.694 & 4.41\\ 
Be$_{12}$Cr & mp-1590 & \multicolumn{2}{c|}{7.205} & 4.113 & \multicolumn{2}{c|}{7.238} & 4.174 & 1.44\\ 
MnAlGe & mp-20757 & \multicolumn{2}{c|}{3.882} & 5.922 & \multicolumn{2}{c|}{3.89} & 5.925 & 2.02\\ 
Fe$_5$B$_2$P & mp-9913 & \multicolumn{2}{c|}{5.441} & 10.31 & \multicolumn{2}{c|}{5.477} & 10.345 & 8.52\\ 
\hline 
CrTe & mp-794 & \multicolumn{2}{c|}{4.128} & 6.273 & \multicolumn{2}{c|}{4.005} & 6.242 & 3.88\\ 
MnBi & mp-568382 & \multicolumn{2}{c|}{4.307} & 5.738 & \multicolumn{2}{c|}{4.305} & 6.118 & 3.52\\ 
Fe$_2$P & mp-778 & \multicolumn{2}{c|}{5.800} & 3.432 & \multicolumn{2}{c|}{5.877} & 3.437 & 2.96\\ 
Fe$_3$N & mp-1804 & \multicolumn{2}{c|}{4.652} & 4.317 & \multicolumn{2}{c|}{4.708} & 4.388 & 6.17\\ 
YCo$_5$ & mp-2827 & \multicolumn{2}{c|}{4.912} & 3.941 & \multicolumn{2}{c|}{4.951} & 3.975 & 7.03\\ 
MnAs & mp-610 & \multicolumn{2}{c|}{3.665} & 5.498 & \multicolumn{2}{c|}{3.710} & 5.691 & 2.90\\ 
MnSb & mp-786 & \multicolumn{2}{c|}{4.094} & 5.602 & \multicolumn{2}{c|}{4.12} & 5.79 & 3.19\\ 
ZrFe$_2$ & mp-1190681 & \multicolumn{2}{c|}{4.970} & 8.076 & \multicolumn{2}{c|}{4.952} & 16.12 & 3.07\\ 
HfFe$_2$ & mp-956096 & \multicolumn{2}{c|}{4.944} & 8.027 & \multicolumn{2}{c|}{4.968} & 8.098 & 3.05\\ 
Fe$_3$Ge & mp-21078 & \multicolumn{2}{c|}{5.129} & 4.210 & \multicolumn{2}{c|}{5.178} & 4.226 & 6.42\\ 
Fe$_3$Sn & mp-20883 & \multicolumn{2}{c|}{5.468} & 4.302 & \multicolumn{2}{c|}{5.458} & 4.361 & 6.99\\ 
YFe$_3$ & mp-1192321 & \multicolumn{2}{c|}{5.087} & 16.20 & \multicolumn{2}{c|}{5.133} & 16.39 & 5.57\\ 
Fe$_5$Si$_3$ & mp-449 & \multicolumn{2}{c|}{6.697} & 4.677 & \multicolumn{2}{c|}{6.541} & 4.558 & 7.62\\ 
Mn$_5$Ge$_3$ & mp-617291 & \multicolumn{2}{c|}{7.121} & 4.948 & \multicolumn{2}{c|}{7.184} & 5.053 & 13.3\\ 
\hline 
MnP & mp-2662 & 3.150 & 5.186 & 5.842 & 3.172 & 5.258 & 5.918 & 1.38\\
FeB & mp-20787 & 2.942 & 3.990 & 5.405 & 2.946 & 4.053 & 5.495 & 1.16 \\
Mn$_3$Sn$_2$ & mp-600428 & 5.398 & 7.457 & 8.420 & 5.510 & 7.565 & 8.598 & 7.74 \\
Fe$_3$B & mp-973682 & 4.365 & 5.403 & 6.643 & 4.387 & 5.437 & 6.71 & 6.08 \\
Fe$_3$C & mp-510623 & 4.478 & 5.033 & 6.722 & 4.512 & 5.082 & 6.742 & 5.58 \\
Co$_3$B & mp-20373 & 4.396 & 5.145 & 6.603 & 4.418 & 5.232 & 6.636 & 3.53 \\
Co$_3$C & mp-20925 & 4.426 & 4.93 & 6.686 & 4.483 & 5.033 & 6.731 & 3.05 \\
GdNi & mp-542632 & 3.784 & 10.40 & 4.211 & 3.778 & 10.36 & 4.221 & 7.39 \\
AlFe$_2$B$_2$ & mp-3805 & 2.913 & 11.02 & 2.847 & 2.926 & 11.02 & 2.871 & 2.63 \\
\hline
\end{tabular}
\label{tab:latcon}
\end{table*}

The tensor $K^\lambda_{\alpha\beta}$ was calculated using the smooth Fourier interpolation technique \cite{Pickett1988} implemented in the BoltzTrap2 package \cite{BoltzTraP2}, which was slightly modified to obtain spin-resolved information. After structural optimization, the band eigenvalues were calculated on a $\Gamma$-centered $k$-point mesh with the distance between the mesh points in the direction of each reciprocal lattice vector set to approximately \SI{0.1}{\per\angstrom}. The smooth Fourier interpolation included approximately 20 times more stars of real-space lattice vectors than irreducible $k$-points. It was checked that these parameters provided results for $\tilde\theta_{ij}$ converged to about 0.01.

\section{Spin Hall angles}
\label{sec:results}

Table \ref{tab:results} lists the calculated values of $\tilde\theta_{ij}=\frac12(P_i-P_j)$
for selected materials. Note that $\tilde\theta_{xy}=0$ by symmetry in tetragonal and hexagonal magnets.

\begin{table*}
\caption{Spin polarizations $P_i$ of the principal values of tensor $K^\lambda_{\alpha\beta}$ and maximal exchange-driven spin Hall angles $\tilde\theta_{ij}$ calculated assuming $P_\tau=0$ and $\theta=\pi/4$ for selected magnetic materials. $T_c$ (K) is the experimental Curie temperature. Substrate/Magnet: a possible substrate, obtained using the Materials Project \cite{MaterialsProject}, and its epitaxial relation, such that the interface normal of the magnetic material is not oriented along any of the principal axes of its $K^\lambda_{\alpha\beta}$ tensor. MCIA: minimal co-incident area (\AA$^2$) for that epitaxy \cite{Zur1984}. $G_{sub}$: crystallographic point group of the substrate. $G_{bil}$ (listed when $G_{sub}=C_{1v}$): crystallographic point group of the bilayer. Substrates allowing a finite macroscopic spin Hall effect for a uniformly magnetized ferromagnetic film ($G_{sub}=C_{1}$ or $G_{sub}=G_{bil}=C_{1v}$) are highlighted in bold. All cases with $G_{sub}=C_{1v}$ and $G_{bil}=C_{1}$ belong to case (3) of Sec. \ref{uniform}. a-TiO$_2$ stands for anatase and TiO$_2$ for rutile.
    }
    \centering
    \begin{tabular}{|l|S[table-format=4.0]|S[table-format=3.3]|S[table-format=3.3]|S[table-format=3.3]|S[table-format=3.3]|S[table-format=3.3]|S[table-format=3.3]|l|c|l|l|}
    \hline
    Material & $T_c$ & $P_x$ & $P_y$ & $P_z$ & $\tilde\theta_{zx}$ & $\tilde\theta_{zy}$ & $\tilde\theta_{xy}$ & Substrate/Magnet & MCIA & $G_{sub}$ & $G_{bil}$\\
    \hline
FePt  & 753 &  \multicolumn{2}{S[table-format=3.3]|}{0.34} &  0.24 & \multicolumn{2}{S[table-format=3.3]|}{-0.05} & \multicolumn{1}{c|}{0} &   \textbf{NdGaO$_3$ (011)/(101)} & 51 & $C_{1v}$ & $C_{1v}$\\
FePd  & 1193 &  \multicolumn{2}{S[table-format=3.3]|}{0.37} &  0.31 & \multicolumn{2}{S[table-format=3.3]|}{-0.03} & \multicolumn{1}{c|}{0} &  \textbf{NdGaO$_3$ (011)/(101)} & 51 & $C_{1v}$ & $C_{1v}$\\
CoPt  & 846 &  \multicolumn{2}{S[table-format=3.3]|}{0.23} &  0.07 & \multicolumn{2}{S[table-format=3.3]|}{-0.08} & \multicolumn{1}{c|}{0} &   \textbf{YAlO$_3$ (011)/(101)} & 51 & $C_{1v}$ & $C_{1v}$\\
MnAl  & 650 &  \multicolumn{2}{S[table-format=3.3]|}{0.05} &  0.55 &  \multicolumn{2}{S[table-format=3.3]|}{0.25} & \multicolumn{1}{c|}{0} &   \textbf{NdGaO$_3$ (011)/(101)} & 49 & $C_{1v}$ & $C_{1v}$\\
MnGa  & 629 &  \multicolumn{2}{S[table-format=3.3]|}{0.07} &  0.64 &  \multicolumn{2}{S[table-format=3.3]|}{0.29} & \multicolumn{1}{c|}{0} &   \textbf{NdGaO$_3$ (011)/(101)} & 49 & $C_{1v}$ & $C_{1v}$\\
FeNi  & 842 &  \multicolumn{2}{S[table-format=3.3]|}{0.27} &  0.36 &  \multicolumn{2}{S[table-format=3.3]|}{0.04} & \multicolumn{1}{c|}{0} &   h-BN (0001)/(101) & 11 & $C_{6v}$ & \\
Fe$_2$B & 1013 &  \multicolumn{2}{S[table-format=3.3]|}{0.37} &  0.06 & \multicolumn{2}{S[table-format=3.3]|}{-0.16} & \multicolumn{1}{c|}{0} &  LiGaO$_2$ (010)/(101) & 33 & $C_{1v}$ & $C_{1}$ \\
Co$_2$B & 433 &  \multicolumn{2}{S[table-format=3.3]|}{0.34} &  0.63 &  \multicolumn{2}{S[table-format=3.3]|}{0.14} & \multicolumn{1}{c|}{0} &    LiGaO$_2$ (010)/(101) & 32 & $C_{1v}$ & $C_{1}$\\
MnAu$_4$ & 385 &  \multicolumn{2}{S[table-format=3.3]|}{0.45} &  0.49 &  \multicolumn{2}{S[table-format=3.3]|}{0.02} & \multicolumn{1}{c|}{0} &         GaSe (0001)/(101) & 51 & $C_{6v}$ & \\
VAu$_4$ & 60 & \multicolumn{2}{S[table-format=3.3]|}{-0.54} & -0.58 & \multicolumn{2}{S[table-format=3.3]|}{-0.02} & \multicolumn{1}{c|}{0} &         & & &\\
Mn$_2$Sb & 580 &  \multicolumn{2}{S[table-format=3.3]|}{0.50} &  0.15 & \multicolumn{2}{S[table-format=3.3]|}{-0.17} & \multicolumn{1}{c|}{0} &    MgF$_2$ (101)/(111)  & 78 & $C_{1v}$ & $C_{1}$\\
Mn$_2$Sb$+U$ & 580 & \multicolumn{2}{S[table-format=3.3]|}{0.35} & -0.09 & \multicolumn{2}{S[table-format=3.3]|}{-0.22} & \multicolumn{1}{c|}{0} &    MgF$_2$ (101)/(111) & 78 & $C_{1v}$ & $C_{1}$\\
Fe$_3$B & 828 &  \multicolumn{2}{S[table-format=3.3]|}{0.47} &  0.70 &  \multicolumn{2}{S[table-format=3.3]|}{0.12} & \multicolumn{1}{c|}{0} &        & & &\\
Fe$_3$P & 693 &  \multicolumn{2}{S[table-format=3.3]|}{0.70} &  0.77 &  \multicolumn{2}{S[table-format=3.3]|}{0.03} & \multicolumn{1}{c|}{0} &  Fe$_2$O$_3$ (0001)/(101) & 91 & $C_{3v}$ &\\
Fe$_8$N & 1500 &  \multicolumn{2}{S[table-format=3.3]|}{0.03} &  0.26 & \multicolumn{2}{S[table-format=3.3]|}{0.12} & \multicolumn{1}{c|}{0} &  BaTiO$_3$ (110)/(101) & 48 & $C_{1v}$ & $C_{1}$\\
Mn$_2$Ga$_5$ & 450 &  \multicolumn{2}{S[table-format=3.3]|}{0.47} &  0.20 & \multicolumn{2}{S[table-format=3.3]|}{-0.14} & \multicolumn{1}{c|}{0} &       TiO$_2$ (100)/(101) & 83 & $C_{2v}$ &\\
Be$_{12}$Cr & 50 & \multicolumn{2}{S[table-format=3.3]|}{-0.13} & -0.54 & \multicolumn{2}{S[table-format=3.3]|}{-0.21} & \multicolumn{1}{c|}{0} &   & & &\\
MnAlGe & 520 &  \multicolumn{2}{S[table-format=3.3]|}{0.42} &  0.88 & \multicolumn{2}{S[table-format=3.3]|}{0.23} & \multicolumn{1}{c|}{0} &    \textbf{a-TiO$_2$ (101)/(101)} & 83 & $C_{1v}$ & $C_{1v}$\\
Fe$_5$B$_2$P & 628 &  \multicolumn{2}{S[table-format=3.3]|}{0.23} &  0.17 & \multicolumn{2}{S[table-format=3.3]|}{-0.03} & \multicolumn{1}{c|}{0} &   GdScO$_3$ (001)/(101) & 64 & $C_{2v}$ &\\
\hline
CrTe  & 340 &  \multicolumn{2}{S[table-format=3.3]|}{0.30} &  0.68 &  \multicolumn{2}{S[table-format=3.3]|}{0.19} & \multicolumn{1}{c|}{0} &    LiAlO$_2$ (110)/(10$\bar1$1) & 90 & $C_{2}$ &\\
MnBi & 633 &  \multicolumn{2}{S[table-format=3.3]|}{0.86} &  0.42 & \multicolumn{2}{S[table-format=3.3]|}{-0.22} & \multicolumn{1}{c|}{0} & o-WTe$_2$ (001)/(10$\bar1$1) & 89 & $C_{2v}$ &\\
Fe$_2$P & 217 &  \multicolumn{2}{S[table-format=3.3]|}{0.28} &  0.43 &  \multicolumn{2}{S[table-format=3.3]|}{0.07} & \multicolumn{1}{c|}{0} &    LiF (110)/(10$\bar1$1) & 71 & $C_{2v}$ &\\
Fe$_3$N & 858 & \multicolumn{2}{S[table-format=3.3]|}{-0.14} & -0.01 &  \multicolumn{2}{S[table-format=3.3]|}{0.07} & \multicolumn{1}{c|}{0} &   C (0001)/(11$\bar2$1) & 79 & $C_{6v}$ &\\
YCo$_5$ & 980 &  \multicolumn{2}{S[table-format=3.3]|}{0.51} &  0.63 &  \multicolumn{2}{S[table-format=3.3]|}{0.06} & \multicolumn{1}{c|}{0} &   GaN (10$\bar1$1)/(10$\bar1$1) & 57  & $C_{1v}$ & $C_{1}$\\
MnAs & 318 &  \multicolumn{2}{S[table-format=3.3]|}{0.75} &  0.29 & \multicolumn{2}{S[table-format=3.3]|}{-0.23} & \multicolumn{1}{c|}{0} &    GaN (10$\bar1$1)/(10$\bar1$1) & 116  & $C_{1v}$ & $C_{1}$\\
MnSb & 851 &  \multicolumn{2}{S[table-format=3.3]|}{0.91} &  0.33 & \multicolumn{2}{S[table-format=3.3]|}{-0.29} & \multicolumn{1}{c|}{0} &    MgF$_2$ (110)/(10$\bar1$1) & 82  & $C_{2v}$ &\\
ZrFe$_2$ & 630 & \multicolumn{2}{S[table-format=3.3]|}{-0.38} & -0.41 & \multicolumn{2}{S[table-format=3.3]|}{-0.02} & \multicolumn{1}{c|}{0} &       & & & \\
HfFe$_2$ & 600 & \multicolumn{2}{S[table-format=3.3]|}{-0.26} & -0.05 &  \multicolumn{2}{S[table-format=3.3]|}{0.10} & \multicolumn{1}{c|}{0} &  MgO (100)/(11$\bar2$1)      & 72  & $C_{4v}$ &\\
Fe$_3$Ge & 670 &  \multicolumn{2}{S[table-format=3.3]|}{0.16} &  0.46 &  \multicolumn{2}{S[table-format=3.3]|}{0.15} & \multicolumn{1}{c|}{0} &  Cu(100)/(10$\bar1$1)       & 94 & $C_{4v}$ &\\
Fe$_3$Sn & 748 &  \multicolumn{2}{S[table-format=3.3]|}{0.72} &  0.52 & \multicolumn{2}{S[table-format=3.3]|}{-0.10} & \multicolumn{1}{c|}{0} &         LaAlO$_3$ (10$\bar1$0)/(10$\bar1$1) & 70 & $C_{1v}$ & $C_{1}$\\
YFe$_3$ & 552 & \multicolumn{2}{S[table-format=3.3]|}{-0.08} & 0.21 &  \multicolumn{2}{S[table-format=3.3]|}{0.15} & \multicolumn{1}{c|}{0} &        & & &\\
Fe$_5$Si$_3$ & 383 &  \multicolumn{2}{S[table-format=3.3]|}{0.51} &  0.20 & \multicolumn{2}{S[table-format=3.3]|}{-0.15} & \multicolumn{1}{c|}{0} &  GaAs (100)/(11$\bar2$1) & 67 & $C_{2v}$ & \\
Mn$_5$Ge$_3$ & 296 &  \multicolumn{2}{S[table-format=3.3]|}{0.54} &  0.66 &  \multicolumn{2}{S[table-format=3.3]|}{0.06} & \multicolumn{1}{c|}{0} &      BaF$_2$ (100)/(11$\bar2$1) & 76 & $C_{4v}$ &\\
\hline
MnP & 293 &  0.27 & -0.23 &  0.24 & -0.01 &  0.24 &  0.25 &    MgO (100)/(110) & 36 & $C_{4v}$ &\\
FeB & 598 & -0.53 & -0.36 &  0.21 &  0.37 &  0.28 & -0.09 &   \textbf{SiO$_2$ (10$\bar1$0)/(110)} & 27 & $C_{1}$ & \\
Mn$_3$Sn$_2$ & 262 & -0.25 &  0.44 &  0.27 &  0.26 & -0.09 & -0.35 & CaF$_2$ (100)/(011) & 61 & $C_{4v}$ &\\
Fe$_3$B & 828 &  0.66 &  0.40 &  0.69 &  0.02 &  0.14 &  0.13 &   MgO (100)/(011) & 37 & $C_{4v}$ &\\
Fe$_3$C & 480 &  0.25 & -0.01 &  0.46 &  0.11 &  0.24 &  0.13 &   \textbf{SiO$_2$ (10$\bar1$0)/(101)} & 81 & $C_{1}$ & \\
Co$_3$B & 747 &  0.66 &  0.52 &  0.73 &  0.04 &  0.11 &  0.07 &  \textbf{ZrO$_2$ (101)/(101)} & 41 & $C_{1v}$ & $C_{1v}$\\
Co$_3$C & 498 &  0.38 &  0.38 &  0.49 &  0.05 &  0.06 &  0.00 &  \textbf{SiO$_2$ (10$\bar1$0)/(101)} & 79 & $C_{1}$ & \\
GdNi & 78 &  0.10 & -0.11 & -0.19 & -0.14 & -0.04 &  0.10 &  BaTiO$_3$ (111)/(101) & 58 & $C_{1v}$ & $C_{1}$\\
AlFe$_2$B$_2$ & 285 &  0.73 &  0.55 &  0.73 & -0.00 &  0.09 &  0.09 &  CdS (0001)/(111) & 46 & $C_{6v}$ &\\
    \hline
    \end{tabular}
\label{tab:results}
\end{table*}

Some of the largest values of $|\tilde\theta_{ij}|$ are found in L1$_0$-ordered MnGa (0.29), hexagonal MnSb (0.29), and orthorhombic FeB (0.37) and Mn$_3$Sn$_2$ (0.35). These values compare favorable with commonly used heavy-metal spin Hall sources, such as Pt and $\beta$-W \cite{Sinova-SHE,Manchon_RMP}.

The last two columns of Table \ref{tab:results} include possible substrates, selected with the help of the Materials Project database \cite{MaterialsProject}, which can form a lattice-matched interface with the given material, and the corresponding minimal co-incident area (MCIA) \cite{Zur1984} for that epitaxy. All of these epitaxial relations have the ferromagnetic film oriented so that there is no principal axis perpendicular to the film plane, thereby allowing spin Hall effect for a given orientation domain. Of course, there is no guarantee that these epitaxial relations can be realized in practice, and they should be regarded as illustrative examples.

\section{Spin Hall domains}
\label{sec:domains}

As will be discussed shortly below, many epitaxial relations allow more than one energetically degenerate orientation domain of the ferromagnetic film on the same substrate, characterized by different spin Hall conductivity tensors. In a large-area film, degenerate orientation domains should occur with equal probabilities, and macroscopic physical responses correspond to statistical averages over these domains. Of course, several nondegenerate interface terminations may coexist in one sample, resulting in more than one set of degenerate orientation domains occurring with different probabilities. This section discusses macroscopic spin Hall response of a multidomain ferromagnetic film from the symmetry point of view. If this response survives after statistical averaging over orientation domains, it can be observed either using macroscopic probes over length scales exceeding the orientation domain size or as an ensemble average over single-domain nanoscale devices where orientations domains occur at random.

I start with general considerations and then consider the case of uniform magnetization in Section \ref{uniform}, which imposes stringent symmetry requirements on the substrate and the epitaxial relation. Then Section \ref{strong-mca} turns to the case of strong magnetocrystalline anisotropy, which enables macroscopic spin Hall response even for high-symmetry substrates.

\subsection{General considerations}

Let $G_{sub}$ and $G_\mathrm{FM}$ be the (nonmagnetic) surface point groups of the substrate alone and the ferromagnetic layer alone, respectively, and $G_{bil}$ the point group of the epitaxial bilayer. $G_{bil}$ is a common subgroup of $G_{sub}$ and $G_\mathrm{FM}$, and I assume $G_{bil}=G_{sub}\cap G_\mathrm{FM}$ when viewed as a set. Of course, $G_{bil}$ depends on the mutual orientation of the symmetry elements of $G_{sub}$ and $G_\mathrm{FM}$. 

Neglecting the non-symmorphic components of the symmetry operations automatically incorporates statistical averaging over interface roughness \cite{boundarymag}. This is the reason to consider point groups rather than space groups.

Because the ferromagnetic layer is assumed to have tilted principal axes of the conductivity tensor, we require that $G_\mathrm{FM}$ is at most $C_{1v}$, which implies that $G_{bil}$ is also at most $C_{1v}$.

If $G_{bil}$ is a \emph{proper} subgroup of $G_{sub}$, the same substrate supports several energetically degenerate crystallographic orientations of the ferromagnetic layer grown on top of that substrate. The set of these degenerate orientations is analogous to the orientation states of a ferroic crystal \cite{Aizu}. Different orientation domains are mapped onto each other by the symmetry operations in $G_{sub}$. The number of inequivalent orientation domains is equal to the index of $G_{bil}$ in $G_{sub}$ (see Theorem 3 of Ref. \cite{Aizu}), and they can be classified by the left cosets of $G_{bil}$ in $G_{sub}$.

\subsection{Uniform magnetization}
\label{uniform}

Here I discuss statistical averaging of the spin Hall conductivity over one set of degenerate orientation domains that are all magnetized in the same direction, allowing the spin degree of freedom to be treated as a scalar quantity. This case is realized when the magnetocrystalline anisotropy is weak, and the entire film is uniformly magnetized by an application of an external magnetic field.

In the most general case, the effective longitudinal conductivities (\ref{sigmaxx}) and the effective Hall conductivities of different orientation domains from the same degenerate set may be different, leading to an inhomogeneous current density and electric field distribution. The inhomogeneity of the electric field over different orientation domains can, in principle, lead to a finite spin Hall response even if the average spin Hall conductivity vanishes. I will disregard this possibility and assume that the effective charge conductivity is approximately isotropic. 

Generalizing Eq. (\ref{sigmaSH}) to an arbitrary direction of the in-plane electric field, the spin Hall current flowing perpendicular to the film can be written as $j^s_z=\mathbf{p}_\mathrm{SH}\mathbf{E}$, where $(\mathbf{p}_\mathrm{SH})_\alpha=\sigma^s_{z\alpha}-\sigma^{-1}_{zz}\sigma^s_{zz}\sigma_{z\alpha}$ and $\alpha\in\{x,y\}$. The in-plane polar vector $\mathbf{p}_\mathrm{SH}$ describes the out-of-plane spin Hall current of a given orientation domain and is invariant under $G_\mathrm{FM}$. In the case considered in the derivation of Eq. (\ref{SHA}), when one principal axis of $\hat \sigma^\lambda$ lies in the plane, $\mathbf{p}_\mathrm{SH}$ is perpendicular to that axis.

The star of vectors $\mathbf{p}^{(i)}_\mathrm{SH}$ describing different orientation domains $i$ belonging to the same degenerate set is generated by the symmetry operations from $G_{sub}$. The macroscopic spin Hall response is given by an average over this star. For it to be finite, it is necessary that $G_{sub}$ allows an in-plane polar vector, which is only the case for $G_{sub}=C_{1}$ or $C_{1v}$. For $G_{sub}=C_{1}$ there is only one orientation domain, and it is sufficient that $\mathbf{p}_\mathrm{SH}\neq0$ for it.

For $G_{sub}=C_{1v}$, let $\sigma_v$ be the mirror plane of the $G_{sub}$ group. There are three possibilities:

\begin{enumerate}
\item $G_{bil}=G_{sub}=C_{1v}$. There is only one orientation domain because $G_{bil}$ is not a proper subgroup of $G_{sub}$. Because we have already assumed that $G_\mathrm{FM}$ is at most $C_{1v}$, and $G_{bil}=C_{1v}$ is its subgroup, it must be that $G_\mathrm{FM}=G_\mathrm{sub}$ with the same $\sigma_v$. Therefore, $\mathbf{p}_\mathrm{SH}\parallel\sigma_v$ and finite.

\item $G_{bil}=C_{1}$ and $\mathbf{p}_\mathrm{SH}$ has a finite projection on $\sigma_v$. This means that either $G_\mathrm{FM}=C_1$ or $G_\mathrm{FM}=C_{1v}$ but its symmetry plane is not orthogonal to $\sigma_v$ of $G_{sub}$. There are two orientation domains but $\mathbf{p}^{(1)}_\mathrm{SH}+\mathbf{p}^{(2)}_\mathrm{SH}\neq0$, and the microscopic average is finite.

\item $G_{bil}=C_{1}$ and $\mathbf{p}_\mathrm{SH}\perp\sigma_v$. This happens if $G_\mathrm{FM}=C_{1v}$ with the symmetry plane orthogonal to $\sigma_v$ of $G_{sub}$. There are two orientation domains with $\mathbf{p}^{(1)}_\mathrm{SH}=-\mathbf{p}^{(2)}_\mathrm{SH}$, and the microscopic average vanishes. 
\end{enumerate}

Thus, a finite macroscopic spin Hall current is allowed for $G_{sub}=C_1$ and only in cases (1) and (2) for $G_{sub}=C_{1v}$. To maximize this current, the electric field should be aligned parallel to the average of $\mathbf{p}^{(i)}_\mathrm{SH}$ over the orientation domains.

\subsubsection{Substrate examples}

The point groups $G_{sub}$ are listed in Table \ref{tab:results} for all substrates.
Many of them are higher than $C_{1v}$, which guarantees that the macroscopic spin Hall conductivity vanishes after averaging over uniformly magnetized degenerate orientation domains. 

A generic orientation of the surface plane would result in $G_{sub}=C_1$ for any crystalline substrate and allow macroscopic spin Hall effect, but this case is uncommon for readily available substrates. In Table \ref{tab:results} it is realized by SiO$_2$ (10$\bar1$0), which may be matched with orthorhombic FeB (110), Fe$_3$C (101), or Co$_3$C (101). The $D_3$ bulk point group of SiO$_2$ has no mirror planes, and the (10$\bar1$0) plane (also known as the $Y$-cut) is not orthogonal to a symmetry axis.

For substrates with $G_{sub}=C_{1v}$, we need to distinguish among the three cases listed above, and Table \ref{tab:results} also includes $G_{bil}$.
Substrates corresponding to Case (1), which allows macroscopic spin Hall effect, have $G_{sub}=G_{bil}=C_{1v}$ and are highlighted in bold in Table \ref{tab:results}. The largest values of $\tilde\theta_{\alpha\beta}$ estimated for these systems are in L1$_0$-ordered MnAl (101) and MnGa (101) on NdGaO$_3$ (011) and for MnAlGe (101) on anatase TiO$_2$ (101) \footnote{The $\theta$ angles corresponding to the epitaxial relations listed in Table \ref{tab:results} are not $\pi/4$, but the effect of this difference is usually minor. For example, for MnAl (101) we have $\theta\approx63^\circ$, and the estimate of $\tilde\theta_{zx}$ is reduced by a factor $\sin2\theta\approx0.81$.}. Note that the L1$_0$-ordered $\tau$-phase of MnAl is metastable but can be stabilized by allowing with Ga \cite{Mix2017}. NdGaO$_3$ (011) and YAlO$_3$ (011) substrates \footnote{The notation for NdGaO$_3$, YAlO$_3$, LiGaO$_2$, and GdSc$O_3$ assumes the orthorhombic axes are ordered by increasing lattice constant.} present an interesting case where the $C_{3v}$ point group of the (111) surface of the parent perovskite structure is broken by the bulk orthorhombic distortion, resulting in $G_{sub}=C_{1v}$.

Case (2) for $G_{sub}=C_{1v}$ requires that the symmetry planes of the substrate and the ferromagnet are neither parallel nor orthogonal to each other. I have not found examples of this case.

Case (3) for $G_{sub}=C_{1v}$, which results in zero macroscopic spin Hall conductivity, is rather common and is realized, for example, in (101)-oriented FePt or MnGa grown on rutile TiO$_2$ (101) and in all other cases in Table \ref{tab:results} with $G_{sub}=C_{1v}$ that are not highlighted.

\subsection{Case of strong magnetocrystalline anisotropy}
\label{strong-mca}

In this case, the magnetizations in different orientation domains are generally not collinear with each other. As I show below, this allows one to obtain a finite macroscopically averaged spin Hall current even if $G_{sub}$ includes a high-order symmetry axis, and the spin polarization of that current can be switched both in sign and in orientation by changing the direction of the initializing external magnetic field and the direction of the current flow.

Because there is no longer a globally defined direction of magnetization, we need to restore the vector character of the spin polarization when averaging over the orientation domains. I will assume that the magnetization is uniform inside each orientation domain, which is reasonable if the domain size is large compared to the magnetic domain wall width. The out-of-plane spin current for orientation domain $i$ can be written as $\mathbf{j}^{(i)}_s=\mathbf{\hat m}^{(i)}(\mathbf{p}^{(i)}_\mathrm{SH}\cdot\mathbf{E})$, where the unit vector $\mathbf{\hat m}^{(i)}$ denotes the direction of the magnetization in orientation domain $i$. The subscript $z$ for the direction of spin current flow has been omitted but is implied throughout. The macroscopic average of the $3\times2$ spin conductivity tensor is given by the tensor $\hat\sigma_\mathrm{SH}=\sum_i\mathbf{\hat m}^{(i)}\otimes\mathbf{p}^{(i)}_\mathrm{SH}/N_i$, where $N_i$ is the number of orientation domains.

To be specific, let us consider the case of dominant bulk easy-axis anisotropy in a tetragonal or hexagonal ferromagnet with $G_\mathrm{FM}=C_{1v}$. In this case the magnetization lies in the plane containing the surface normal and the bulk fourfold or sixfold (or threefold) axis, which, as above, is tilted by angle $\theta$ with respect to the $z$ axis. The angle made by $\mathbf{\hat m}^{(i)}$ with the $z$ axis will be denoted as $\theta_m$, which may differ from $\theta$ due to the demagnetizing field but is the same for all $i$, at least once the external field has been switched off.

Along with the $\mathbf{p}^{(i)}_\mathrm{SH}$ vector, the operations in $G_{sub}$ also transform the orientation of the easy magnetization axis while keeping $\theta$ unchanged. First, assume the sample is magnetized by an out-of-plane magnetic field $\mathbf{B}=B_z\hat z$, so that $m^{(i)}_z$ is the same in all orientation domains. Because $m^{(i)}_z$ behaves as a scalar under $G_{sub}$, the $z$-polarized spin current behaves exactly as described in Section \ref{uniform} for the case of uniform magnetization.

On the other hand, the in-plane projection $\mathbf{m}^{(i)}_\parallel$ transforms like a polar vector under $G_{sub}$, similar to $\mathbf{p}^{(i)}_\mathrm{SH}$. (That $\mathbf{m}^{(i)}_\parallel$ formally behaves here as a polar rather than axial vector is enforced by the external field which selects the same sign of $m^{(i)}_z$ for all orientation domains.) Our assumption of $G_\mathrm{FM}=C_{1v}$ leads to $\mathbf{m}^{(i)}_\parallel\parallel\mathbf{p}^{(i)}_\mathrm{SH}$. Thus, the spin Hall conductivity $\hat\sigma^\parallel_\mathrm{SH}$ giving the out-of-plane flow of in-plane-polarized spins is equal to the average of $\pm p_\mathrm{SH}\,\hat{p}^{(i)}_\mathrm{SH}\otimes\hat{p}^{(i)}_\mathrm{SH}\sin\theta_m$ over all orientation domains, where $\hat{p}^{(i)}_\mathrm{SH}=\mathbf{p}^{(i)}_\mathrm{SH}/p_\mathrm{SH}$ and the overall sign is that of $\mathbf{m}^{(i)}_\parallel\cdot\mathbf{p}^{(i)}_\mathrm{SH}$, which in turn depends on the sign of $B_z$. The average over $i$ survives under \emph{any} point group $G_{sub}$. If $G_{sub}$ includes a threefold or higher rotation axis, $\hat\sigma^\parallel_\mathrm{SH}$ reduces to a scalar $\sigma^\parallel_\mathrm{SH}=\pm\frac12 p_\mathrm{SH}\sin\theta_m$. Thus, we find an interesting situation where the spin polarization of the macroscopically averaged spin current flowing in the $z$ direction is locked parallel to the direction of the in-plane charge current flow. This is in contrast to the conventional spin-orbit-driven spin Hall effect in nonmagnetic materials where the spin current would be polarized perpendicular to the current flow direction.

The situation is more complicated if the system is initialized by an in-plane magnetic field $\mathbf{B}$, rather than out-of-plane as above. The spin Hall current depends on $G_{sub}$, on the orientation of the initializing magnetic field, and on the orientation of the charge current flow, with respect to the symmetry elements of $G_{sub}$ and $G_\mathrm{FM}$. For example, consider $G_{sub}=C_{4v}$ and $G_\mathrm{FM}=C_{1v}$, and align the $x$ axis parallel to the coincident mirror planes of $G_{sub}$ and $G_\mathrm{FM}$ for one of the orientation domains. There are four orientation domains with $\mathbf{p}^{(i)}_\mathrm{SH}$ oriented along the positive and negative directions of the $x$ and $y$ axes. For each domain, it is assumed the magnetization picks one of the two directions along its easy axis such that $\mathbf{m}^{(i)}\mathbf{B}>0$. Then the averaged spin Hall current is strictly $z$-polarized: $\mathbf{j}_s=\frac12 \hat z\, p_\mathrm{SH}\cos\theta_m(\pm1,\pm1)\cdot\mathbf{E}$, where the signs for the vector components in brackets are such that this vector lies in the same quadrant with $\mathbf{B}$.

Another interesting example is when $\mathbf{B}$ is applied in the $x$ direction with a small tilt toward $z$. In this case, the average spin Hall current is $\mathbf{j}_s=\frac12 p_\mathrm{SH}(0,E_y\sin\theta_m,E_x\cos\theta_m)$, and its polarization can be switched between in-plane and out-of-plane by changing the direction of the current flow. 

As a function of the orientation of the initializing magnetic field $\mathbf{B}$, the spin Hall conductivity tensor has discontinuities along the contours on the sphere where one or more orientation domains switch their magnetization. After initialization, the measurement may be performed in the remanent state.

\section{Summary}
\label{sec:conclusions}

A crystalline ferromagnetic (or ferrimagnetic) film with tilted principal axes of the conductivity tensor can exhibit exchange-driven spin Hall effect, generating spin-polarized current perpendicular to the film surface, thanks to the anisotropy of the conductivity tensor. This effect is analogous to the spin-splitting current in altermagnets and does not require spin-orbit coupling. First-principles screening of 41 tetragonal, hexagonal, and orthorhombic ferromagnets revealed that, for certain crystallographic orientations, the spin Hall angle is greater than 0.1 in 26 of them, and greater than 0.2 in 12.

To allow the use of this exchange-driven spin Hall effect on a macroscopic scale, the response should not vanish after averaging over the crystallographic orientation domains of the ferromagnetic film. For a uniformly magnetized sample, this is possible only if the substrate has low enough symmetry ($C_1$ or $C_{1v}$); moreover, for a $C_{1v}$ substrate, the ferromagnetic film should not be aligned in such a way that two degenerate orientation domains have opposite spin Hall angles. If these conditions are not met, the macroscopic spin Hall conductivity of a uniformly magnetized film vanishes after averaging over degenerate orientation domains. Several cases listed in Table \ref{tab:results} do meet these conditions, most notably MnAl (101) and MnGa (101) on NdGaO$_3$ (011) and MnAlGe (101) on anatase TiO$_2$ (101), with rather large estimated spin Hall angles. In all systems listed in Table \ref{tab:results} the spin Hall effect may be observed locally for each orientation domain, for example, with a spatially resolved measurement of the magneto-optic Kerr effect.

The spin Hall effect can also be observed on the macroscopic scale, \emph{regardless} of the substrate symmetry, if the ferromagnetic film has magnetocrystalling anisotropy strong enough to lock the magnetization to the crystallographic easy axis in each orientation domain. Multiple possibilities exist depending on the crystallographic symmetry and the orientation of the initializing magnetic field, including the averaged spin Hall current being polarized parallel to the current flow or perpendicular to the film plane.

Exchange-driven spin Hall effect in ferromagnets expands the range of available spin Hall sources for applications in magnetic heterostructures. In particular, one can envision a trilayer heterostructure in which the ferromagnetic source layer is separated from the detector layer by a nonmagnetic spacer, such as Cu, which is sufficiently thick to eliminate the exchange coupling between the layers. If the source ferromagnet has strong magnetocrystalline anisotropy, the spin Hall current can have a large out-of-plane component, facilitating the switching of the detector layer with perpendicular magnetization; this opportunity is advantageous for applications in magnetic random-access memory. In contrast to an altermagnet, the source ferromagnetic layer can be deterministically initialized by an external magnetic field. In terms of its device structure, the ferromagnetic trilayer spin Hall device is similar to a spin-orbit torque bilayer \cite{Manchon_RMP}, where, in contrast with spin-transfer torque devices, the read and write current paths are separated.

\begin{acknowledgments}
I thank Ilya Krivorotov for fruitful discussions and Mark Stiles, Evgeny Tsymbal, Igor Mazin, Vivek Amin, Alexey Kovalev, and Xin Fan for useful comments about the manuscript. This work was supported by the National Science Foundation through Grants No. DMR-1916275 and DMR-2324203. Calculations were performed utilizing the Holland Computing Center of the University of Nebraska, which receives support from the Nebraska Research Initiative.
\end{acknowledgments}

\bibliography{refs}

\end{document}